\documentclass[prl,twocolumn,preprintnumbers,amsmath,amssymb,superscriptaddress,showpacs,longbibliography]{revtex4-1}
\usepackage{graphicx}
\usepackage{latexsym}
\usepackage{amsmath}
\usepackage{graphics}
\usepackage{amssymb}
\usepackage{layout}
\usepackage{verbatim}
\usepackage{amsfonts,epsfig,dsfont}
\usepackage{natbib}
\usepackage{hyperref}

\newcommand{\ket}[1]{\left|#1\right>}
\newcommand{\bra}[1]{\left< #1 \right|}
\newcommand{\beq}{\begin{equation}}
\newcommand{\eeq}{\end{equation}}
\newcommand{\bea}{\begin{eqnarray}}
\newcommand{\eea}{\end{eqnarray}}

\newcommand{\tr}{\hbox{Tr}}

\begin{document}

\title{Characterization and measurement 
of qubit-environment entanglement generation during pure dephasing}

\author{Katarzyna Roszak}
\affiliation{Department of Theoretical Physics, Wroc{\l}aw University of Technology,
50-370 Wroc{\l}aw, Poland}
\affiliation{Department of Condensed Matter Physics, Faculty of Mathematics and Physics, 
Charles University, 12116 Prague, Czech Republic}

\author{{\L}ukasz Cywi{\'n}ski}
\affiliation{Institute of Physics, Polish Academy of Sciences, 02-668 Warsaw, Poland}

\date{\today}

\begin{abstract}
We consider the coupling of a qubit in a pure state to an environment in an arbitrary state, and characterize the possibility of qubit-environment entanglement generation during the evolution of the joint system, that leads to pure dephasing of the qubit. We give a simple necessary and sufficient condition on the initial density matrix of the environment together with the properties of the interaction, for appearance of qubit-environment entanglement. Any entanglement created
turns out to be detectable by the Peres-Horodecki criterion. Furthermore, we show that for a large family of initial environmental states, the appearance of nonzero entanglement with the environment is necessarily accompanied by a change in the state of the environment (i.e.~by the back-action of the qubit). 
\end{abstract}

\maketitle

When a quantum system, a qubit (Q) in the context of this paper, is coupled to an environment (E), an initial pure state of the qubit evolves into a mixed state. It is widely recognized that there is an intimate relation between this process of decoherence \cite{Zurek_RMP03,Hornberger_LNP09} and creation of qubit-environment entanglement (QEE). A precise statement can be made for E initialized in a pure state: the establishment of QEE is then equivalent to the reduction of purity of the reduced density matrix of Q. However, in the case of an initial mixed state of E, the situation is more complicated: the state of Q can lose its purity while no QEE is established \cite{Eisert_PRL02}. This should not be surprising, since QEE should be associated with decoherence, understood as a process in which information is transferred from Q to E (i.e.~E is in some sense ``measuring'' Q), not with the bare fact that the state of Q is becoming mixed. The latter can happen when there is \textit{no influence} of Q on E. For example, when the self-Hamiltonian $\hat{H}_{\text{E}}$ of E commutes with the qubit-environment interaction $\hat{V}_{\text{QE}}$, and when the initial state of E fulfills  $\hat{\rho}_{\text{E}}(0) \!= \! f(\hat{H}_{\text{E}})$ (e.g.~it is a thermal state determined by $\hat{H}_{\text{E}}$), there is no back-action of Q on E, while E is simply a source of random, classical, and quasi-static fields acting on Q \cite{Helm_PRA09,Helm_PRA10,LoFranco_PRA12,Liu_SR12}. In this case of so-called random unitary (RU) evolution, the purity of Q decays while no QEE is established. 

A natural question to ask is whether the RU case is the only one for which QEE does not accompany the loss of purity of Q. While specific kinds of environments and  qubit-environment couplings were investigated in this context (e.g.~quantum Brownian motion \cite{Eisert_PRL02,Hilt_PRA09} or pure dephasing due to a bath of noninteracting bosons \cite{Pernice_PRA11}), the general answer to this question seems to be lacking. This is to a large degree caused by the fact that quantification (or even checking for presence) of QEE is a very hard problem when mixed states of the total system are considered, and when the dimension of Hilbert space of E is larger than $3$ \cite{Mintert_PR05,Plenio_QIC07,Horodecki_RMP09,Aolita_RPP15}. The existence of bound entanglemet \cite{Horodecki_PLA97,Horodecki_PRL98} (which is not detected by the Peres-Horodecki criterion \cite{Peres_PRL96,Horodecki_PLA96} of negativity of a partially transposed matrix of the total system) severely hampers the task of general understanding of QEE \cite{Kraus_PRA00}. 

In this paper we present a complete characterization of conditions for the appearance of QEE in a less general, but physically well-motivated situation. Specifically, we focus on the case of pure dephasing (PD) of the qubit, defined by condition that $[\hat{H}_{\text{Q}},\hat{V}_{\text{QE}}] \! = \! 0$ (where $\hat{H}_{\text{Q}}$ is the self-Hamiltonian of Q). The PD case is not only the paradigmatic example for relation between decoherence of Q and establishment of QEE \cite{Zurek_RMP03}, it is also often encountered in experiments, when the energy splitting of Q is tuned to values for which the exchange of energy between Q and E is suppressed either because of diminished coupling,
or because of diminished density of states of E with matching energies. 
In such a constrained setup we derive the necessary and sufficient condition for nonzero QEE. This condition is that for the initial state of the environment $\hat{\rho}_{\text{E}}(0) \! =\! \sum_{n} c_{n} \ket{n}\bra{n}$, QEE is absent
at time $t$ iff for any states $\ket{i}$ and $\ket{n}$ for which $c_{i} \! \neq \! c_{n}$ the qubit-induced evolution evolution of E \textit{does not couple these states} (see below for a more precise formulation). This condition shows the lack of QEE for a completely mixed $\hat{\rho}_{\text{E}}(0)$
(this statement is not trivial, as one may think, 
since there exist non-pure-dephasing evolutions that lead to QEE when the initial density matrix of the environment proportional to unity, see Appendix for an example), for the RU case, and also for a family of cases which correspond to a mixture of the former two (i.e.~when both $\hat{\rho}_{\text{E}}(0)$ and $\hat{V}_{\text{QE}}$ are nontrivially constrained). Furthermore, QEE is always detected by the Peres-Horodecki criterion, i.e.~only ``free'' entanglement can be created in the PD process. 

Finally, we show that for a physically well-motivated family of $\hat{\rho}_{\text{E}}(0)$ states, generation of nonzero QEE at time $t$ occurs iff the reduced state of E changes: $\hat{\rho}_{\text{E}}(t) \! \neq \! \hat{\rho}_{\text{E}}(0)$. Thus, measurement of change of \textit{any} environmental observable can be a witness of creation of QEE. This is an interesting example how the common lore of the impossibility of bipartite entanglement detection by measuring one party only, turns out not to hold when some prior knowledge on the initial state and the character of inter-system coupling is available.



We begin with the most general form of the Hamiltonian of Q and E which describes the PD case:
\begin{eqnarray}
\hat{H}&=&
\hat{H}_{\mathrm{Q}}+\hat{H}_{\mathrm{E}}+ |0\rangle\langle 0|\otimes{V_0} +|1\rangle\langle 1|\otimes{V_1} \label{eq:Hgen} \,\, .
\end{eqnarray}
The first term of the Hamiltonian describes the qubit and is given by $H_{\mathrm{Q}}=\sum_{i=0,1}\varepsilon_{i}|i\rangle\langle i|$,  the second  describes the environment, while the remaining terms describe the qubit-environment interaction with the qubit states written on the left side of each term (the environment operators $V_0$ and $V_1$ are arbitrary).

Since entanglement between two subsystems is unaffected by local unitary operations on either of the subsystems \cite{Mintert_PR05,Plenio_QIC07,Horodecki_RMP09,Aolita_RPP15}, the full qubit-environment evolution operator $\hat{U}(t)\! =\! \exp(-i\hat{H}t)$ resulting from the Hamiltonian (\ref{eq:Hgen}) may be transformed into
\begin{equation}
\label{u}
\tilde{U}(t)=e^{i(\hat{H}_{\text{Q}}+\hat{H}_{0} )t}\hat{U}(t) = |0\rangle\langle 0|\otimes\mathds{1}_{\mathrm{E}}+ |1\rangle\langle 1|\otimes \hat{w}(t) \,\, , 
\end{equation}
where $\hat{H}_{i} \! =\! \hat{H}_{\mathrm{E}} + \hat{V}_{i}$ and we have defined the operator 
\begin{equation}
\label{w}
\hat{w}(t)= \exp(i\hat{H}_{0}t)\exp(-i\hat{H}_{1}t) \,\, . 
\end{equation}
Note that while $\hat{H}_{\mathrm{Q}}$ commutes with all the other terms in $\hat{H}$, this is not necessarily the case with $\hat{H}_{\mathrm{E}}$.

To find out what the conditions for the generation of QEE via pure dephasing are, let us study the joint state of a qubit and an environment which are initially in a product state $\hat{\sigma}(0)=\hat{\rho}_{\mathrm{Q}}(0)\otimes\hat{\rho}_{\mathrm{E}}(0)$ and evolve according to the operator (\ref{u}).
The qubit is initially in a pure state $|\psi\rangle=a|0\rangle+b|1\rangle$, with
$|a|^2+|b|^2=1$ and $a,b\neq 0$ (a superposition is needed for dephasing to occur), 
so the density matrix $\hat{\rho}_{\mathrm{Q}}(0)=|\psi\rangle\langle\psi |$ (the issue of QEE becomes more complicated when initial purity of Q is not maximal \cite{Pernice_PRA11}, and we exclude this case here).
At this stage we impose no restrictions on the initial density matrix of the environment and write it in terms of its  eigenstates, $\hat{\rho}_{\mathrm{E}}(0)=\sum_n c_n|n\rangle\langle n|$. The time-evolved qubit-environment density matrix
in the evolution picture described by the operator (\ref{u}) takes the form
\begin{equation}
\label{mac1}
\tilde{\sigma}(t)=\left(
\begin{array}{cc}
|a|^2\sum_n c_n|n\rangle\langle n|&ab^*\sum_n c_n|n\rangle\langle n'(t)|\\
a^*b\sum_n c_n|n'(t)\rangle\langle n|&|b|^2\sum_n c_n|n'(t)\rangle\langle n'(t)|
\end{array}\right),
\end{equation}
where the matrix is written in the basis of the eigenstates of Q, and $|n'(t)\rangle=\hat{w}(t)|n\rangle$ with $\hat{w}(t)$ given by Eq.~(\ref{w}).

Firstly, let us study the simplest situation when the initial state of the environment is completely mixed
(i.e.~$\hat{\rho}_{\mathrm{E}}(0) \! =\! \mathds{1}/N$ where $N$ is the dimension of the environment). 
Using Eqs.~(\ref{u}) and (\ref{mac1}) we arrive at
\begin{equation}
\tilde{\sigma}(t)=\frac{1}{N}\left(
\begin{array}{cc}
|a|^2\mathds{1}&ab^*\mathds{1}\hat{w}^{\dagger}(t)\\
a^*b\hat{w}(t)\mathds{1}&|b|^2\mathds{1}
\end{array}\right).
\end{equation}
Now we use the complete mixedness of E: the unity can be written as a sum of projectors over any basis set, including the basis spanned by eigenvectors of a Hermitian operator $\hat{h}(t)$ defined by $\hat{w}(t) \! \equiv \! \exp[-i\hat{h}(t)]$. With $\hat{h}(t)\ket{k(t)} \! =\! \chi_{k}(t)\ket{k(t)}$ (note that the eigenvectors and eigenvalues evolve with time), we have
\beq
{\tilde{\sigma}}(t) =\frac{1}{N}
\sum_{k}\left(
\begin{array}{cc}
|a|^2&ab^* e^{i\chi_{k}(t)}\\
a^*b e^{-i\chi_{k}(t)}&|b|^2
\end{array}\right)\otimes|k(t)\rangle\langle k(t)|
\eeq
which is by definition separable \cite{Werner_PRA89} and no QEE is created, even though the whole system can evolve in a complicated manner.

In the general case, when $\hat{\rho}_{\text{E}}(0)$  is arbitrary,
the question of QEE cannot be resolved so simply. The Peres-Horodecki criterion (PHC) \cite{Peres_PRL96,Horodecki_PLA96} of negativity of partially-transposed  $\tilde{\sigma}(t)$ detects some, but not all entangled states - the bound-entangled states \cite{Horodecki_PLA97,Horodecki_PRL98}, which have a positive partial transpose, are much harder to detect, even for a constrained case of a system consisting of a qubit and an $N$-dimensional environment \cite{Kraus_PRA00}. We proceed now to check for the existence of negative partial transpose QEE. Later it will turn out that states not shown to be entangled by PHC are in fact separable, i.e.~the bound QEE is never created in the process of pure dephasing of the qubit.

The positive semidefinite matrix has only nonnegative eigenvalues, or equivalently all of its principal minors are non-negative. Checking for the second condition turns out to be manageable for the partial transpose of the density matrix from Eq.~(\ref{mac1}). First we treat the case of $\tilde{\sigma}(t)$ which is full-rank (i.e.~all $c_n\neq 0$).  A class of principal minors can be obtained by symmetrically crossing out $(N-1)$ rows and columns in such a way that only one row and column containing a diagonal density matrix element proportional to $|b|^2$ is left. They are given by
\begin{equation}
\label{macierz}
M_{i}=\text{det} \! \left(
\begin{array}{ccccc}
|a|^2c_0&\cdots&0&a^{*}bc_{i}y^{*}_{i0}\\
\vdots&\ddots&\vdots&\vdots\\
0&\cdots&|a|^2c_n&a^*bc_{i}y^{*}_{iN-1}\\
ab^{*}c_iy_{i0}
&\cdots&ab^{*}c_{i}y_{iN-1}&|b|^2\sum_nc_n|y_{ni}|^2
\end{array}
\right),
\end{equation}
where $i=0,1,\ldots,N-1$ and  $y_{ni} \equiv \langle n|\hat{w}^{\dagger}(t)|i\rangle$.
A simple calculation leads to
\begin{equation}
M_{i}  = |a|^{2N}|b|^2 \left(\prod_{k }c_k\right) \sum_{n=0}^{N-1} \left( c_{n}|y_{ni}|^2 - \frac{c^{2}_{i}}{c_{n}} |y_{in}|^2 \right )   \,\, .\label{eq:MiLR}
\end{equation}
When all $c_{i}$ are equal [$\hat{\rho}_{E}(0) \propto \mathds{1}$ and entanglement is not generated
as shown above], all $M_{i} \! =\! 0$ (this is obtained after noticing that both $\sum_{n} |y_{ni}|^2 \! =\! 1$ and $\sum_{n} |y_{in}|^2 \! =\! 1$). Let us assume now that not all $c_{n}$ are the same. We choose then the minor $M_{i}$ corresponding to the largest $c_{i}$, so that for all $j$ we have $c_{i} \geq c_{j}$ and at least for one $j$ the inequality is sharp. We have then $L_{i} = \sum_{n} c_{n}|y_{ni}|^2 \leq c_{i} \sum_{n} |y_{ni}|^2 = c_{i}$,
while $P_{i} = \sum_{n} \frac{c^{2}_{i}}{c_{n}} |y_{in}|^2 > c_{i} \sum_{n} |y_{in}|^2 = c_{i}$,
in which we assumed that $|y_{ij}| \neq 1$ for $j$ corresponding to at least one $c_{j}$ for which $c_{j}<c_{i}$. In such a case we see that $L_{i} < P_{i}$ and consequently $M_{i}\! < \! 0$ signifying the presence of non-bound (free) entanglement between Q and E. 

In the opposite situation, when $y_{ij} \! =\! 0$ for each $j$ corresponding to $c_{j} \!  < \! c_{i}$ we repeat the same reasoning for the $M_{k}$ minor corresponding to the second-largest $c_{k}$. Then either we discover that the state is entangled, or $y_{kj} \! = \! 0$ for all $j$ such that $c_{j} < c_{k}$. If iterating this procedure fails to sense entanglement (i.e.~if no minors are shown to be necessarily negative), then it means that for any pair of $i$ and $j$ with $c_{i} \! > \! c_{j}$ we have $y_{ij} \! =\! 0$, 
which is equivalent to $\bra{i}\hat{w}^{\dagger}\ket{j} = \bra{j}\hat{w}\ket{i} = 0$,
so the evolution due to qubit-environment coupling does not lead out of the subspace corresponding to $c_{i}$. 
In fact, due to unitarity of $\hat{w}$, the non-negativity of all the $M_{i}$ minors is equivalent to $\hat{w}$ having no matrix elements between states corresponding to different occupations $c_{n}$.
The evolution for which $|n\rangle\rightarrow e^{i\phi_n}|n\rangle$ is a special case of this type when the subspace in which the evolution takes place is limited to the single state $|n\rangle$.

Let us show now that any evolution of this type is nonentangling. If we denote the subspaces by an additional index $s$ in such a way that any state $|n_s\rangle$ belongs to that  subspace and all factors $c_{n_s}=c_s$ are equal, the qubit-environment density matrix from Eq.~(\ref{mac1}) can be rewritten as
\begin{equation}
\label{macsep}
\tilde{\sigma}(t)=\sum_s c_s\left(
\begin{array}{cc}
|a|^2\sum_{n_s} |n_s\rangle\langle n_s|&ab^*\sum_{n_s} |n_s\rangle\langle n_s'(t)|\\
a^*b\sum_{n_s} |n_s'(t)\rangle\langle n_s|&|b|^2\sum_{n_s} |n_s'(t)\rangle\langle n_s'(t)|
\end{array}\right).
\end{equation}
Since the subspaces do not overlap, each $\sum_{n_s} |n_s\rangle\langle n_s|$
is a unity in its own subspace (hence, also $\sum_{n_s} |n_s'(t)\rangle\langle n_s'(t)|$
is the same unity), so the sum over $s$ is a sum of density matrices (up to a normalization)
in which the restricted environment has an initial density matrix proportional to unity.
It has been shown previously that such density matrices remain separable during PD evolution.
Hence each of the matrices can be written in the form $\sum_{p_s}q_{p_s}\hat{\rho}_{p_s}(t)\otimes \hat{R}_{p_s}(t)$
and we get
\begin{equation}
\tilde{\sigma}(t)=\sum_s c_s\sum_{p_s}q_{p_s}\hat{\rho}_{p_s}(t)\otimes \hat{R}_{p_s}(t)=\sum_k p_k\hat{\rho}_{k}(t)\otimes \hat{R}_{k}(t),
\end{equation}
with $p_k=c_s q_{p_s}$ and $k$ numbering all the terms that are part of the summation
(meaning that a single index $k$ corresponds to a unique combination of the indices $s$ and $p_s$).
This shows that given the above conditions, the density matrix $\tilde{\sigma}(t)$ is separable.
The result concurs with the fact that entanglement between
an $N$ dimensional and an $M$ dimensional system will always be free 
entanglement, if the number of eigenvalues of the joint density matrix is smaller or equal to $\max(N,M)$ \cite{Horodecki_PRA00}.

To complete the analysis of QEE generation via PD processes, let us study the situation when $\hat{\rho}_{\mathrm{E}}(0)$ has eigenvalues equal to zero (it is not full rank). Obviously, if all states $|p\rangle$ corresponding  to $c_p=0$ are decoupled from all the other states, meaning that for all $n$ for which $c_n \! \neq \! 0$ and for all $p$ for which $c_p \! = \! 0$, $\langle n|\hat{w}^{\dagger}(t)|p\rangle \! = \!0$, the Hilbert space of E can be reduced in such a way that it no longer contains the set of states $\{|p\rangle\}$. The above procedure for finding entanglement can then be used for such a reduced environment and the same conclusions for when the evolution generates QEE hold.
Let us now study the situation when there exists one state $|p\rangle$ for which $c_p=0$ but 
at least one
$\langle n|\hat{w}^{\dagger}(t)|p\rangle=y_{np}\neq 0$ exists. The minors from Eq.~(\ref{eq:MiLR})
are now given by
\begin{equation}
\label{min1}
M_{i}  = -|a|^{2N}|b|^2 \left(\prod_{k \neq p}c_k\right)  c^{2}_{i}|y_{ip}|^2,
\end{equation}
for $i\neq p$ and $M_{p}\! = \! 0$. Hence, $M_{i}$ corresponding to the non-zero value of $y_{ip}$ is negative and free entanglement is generated.

If there exist $K\ge 2$ environment states $|p\rangle$ for which $c_p=0$, with at least one $y_{np}\neq 0$ for each, all the $M_{i}$ given by Eq.~(\ref{eq:MiLR}) are equal to zero and they are no longer a good subset of minors for  the study of QEE generation, yet there exists $K$ new subsets, which can be used to probe QEE (the subsets are equivalent in the  sense that each is sufficient to detect entanglement).
These are obtained by symmetrically crossing out the same $K-1$ rows and columns from each of the matrices given by Eq.~(\ref{macierz}) in such a way that only one diagonal element equal to zero is left. If the state corresponding to this diagonal element is labeled as $|r\rangle$, the minors are given by 
\begin{equation}
\label{min2}
\tilde{M}_{i}  = -|a|^{2(N-K+1)}|b|^2 \left(\prod_{k \neq r}c_k\right)  c^{2}_{i}|y_{ir}|^2.
\end{equation}
As in the case of a single $c_p=0$, the minor $\tilde{M}_{i}$ corresponding to the non-zero value of $y_{ir}$ is negative and QEE is generated.

This closes the part of the article where we characterize the generation of QEE as a result of the general 
pure dephasing Hamiltonian (\ref{eq:Hgen}). We have shown that the appearance of QEE at a given time depends only on the initial state of the environment, $\hat{\rho}_{\text{E}}(0)$, and on the form of the operator $\hat{w}(t)$ given by Eq.~(\ref{w}). Furthermore the qubit-environment state is separable at time $t$ iff for all eigenstates $\ket{n}$ and $\ket{i}$ of $\hat{\rho}_{\text{E}}(0)$ corresponding to eigenvalues $c_{i} \! \neq \! c_{n}$, we have $\bra{n}\hat{w}(t)\ket{i} \! =\! 0$.
This statement is equivalent to the statement that the qubit-environment state is separable at time $t$ iff 
\beq
[\rho_{\mathrm{E}}(0),\hat{w}(t)] \! = \!0 \, \, . \label{eq:main}
\eeq
This is the main technical result of the first part of the paper.

The above criterion of course confirms the lack of QEE when $\hat{\rho}_{\mathrm{E}}(0) \! \propto \! \mathds{1}$, since the unit operator commutes with any $\hat{w}(t)$. In the RU case the assumptions that $[\hat{H}_{\mathrm{E}},\hat{V}_{i}] \! = \! 0$ and $\hat{\rho}_{\mathrm{E}} \! =\! f(\hat{H}_{\mathrm{E}})$ immediately lead to Eq.~(\ref{eq:main}), with  $\hat{w}(t)|n\rangle=e^{i\phi_n t}|n\rangle$ (this is the case of a quasi-static E leading to inhomogeneous broadening of the energy splitting of Q).
However, it is crucial to note that these examples do not exhaust the $\{\rho_{\mathrm{E}}(0),\hat{w}(t)\}$ pairs that satisfy Eq.~(\ref{eq:main}). This condition is fulfilled when $\rho_{\mathrm{E}}(0)$ has subspaces of equal $c_{n}$, and $\hat{w}(t)$ operator is block-diagonal with respect to these subspaces. This means that qubit-induced evolution can be nontrivial within these subspaces in the absence of QEE. For example, one could take a system with $\hat{\rho}_{\mathrm{E}}(0) \! \propto \! \mathds{1}$, and then by appropriate measurements on E post-select its states in such a way, that a subspace with altered $c_{n}$ is singled out. If the qubit-induced evolution due to $\hat{w}(t)$ does not couple this subspace with the rest of the E states, QEE remains absent, while $\hat{\rho}_{\mathrm{E}}(0)$ is nontrivial and nothing is assumed about commutation of $\hat{H}_{\mathrm{E}}$ and $\hat{V}_{i}$ (except for the condition that there is a subspace closed with respect to evolution generated by $\hat{w}(t)$).   


It is easy to see now that the transformed reduced density matrix of E,
\beq
\tilde{\rho}_{\mathrm{E}}(t) = \tr_{\mathrm{Q}}\tilde{\sigma}(t) = |a|^{2}\hat{\rho}_{\mathrm{E}}(0) + |b|^{2} \hat{w}(t) \hat{\rho}_{\mathrm{E}}(0) \hat{w}^{\dagger}(t) \,\, ,
\eeq
remains constant for any $a$ and $b$, iff the condition of separability from Eq.~(\ref{eq:main}) holds: the lack of QEE is equivalent to $\tilde{\rho}_{\mathrm{E}}(t) \! = \! \hat{\rho}_{\mathrm{E}}(0)$. After transforming back to the ``laboratory frame'' using Eq.~(\ref{u}) we see that when there is no QEE, the actual reduced state of E, $\hat{\rho}_{\mathrm{E}}(t) \! = \! \exp(-i\hat{H}_{0}t) \hat{\rho}_{\mathrm{E}}(0) \exp(i\hat{H}_{0}t)$, will remain unchanged iff $[H_{0},\hat{\rho}_{\mathrm{E}}(0)] \! = \!0$. Our second main result is thus the following: if this commutation condition holds, observation of any change in the reduced state of E during the evolution of the coupled qubit-environment system signifies the presence of QEE (note that all the above derivations could be repeated with $\hat{H}_{1}$ replacing $\hat{H}_{0}$ in Eq.~(\ref{u}) and the two $\hat{H}_{i}$ exchanging places in Eq.~(\ref{w}), leading to analogous condition involving $\hat{H}_{1}$).
In this case generation of entanglement can be verified by measurements performed only on one of the entangled subsystems (namely the environment), which is much more convenient than any viable joint qubit-environment measurements. Let us stress that in general (even when entanglement between two qubits is studied), without any prior information about the initial state and the Hamiltonian, probing the existence of entanglement via measurements on a single subsystem is not possible.
Furthermore, in the discussed case  the detection of \textit{any} change  in the state of E is equivalent to the detection of QEE.

The constraint imposed on $\hat{\rho}_{\mathrm{E}}(0)$ that enables the study of QEE via measurements on E only is not a very restrictive one. It means that $H_{i}$ (with $i \! =\! 0$ or $1$) can only have off-diagonal elements [when it is written in the eigenbasis of $\hat{\rho}_{\mathrm{E}}(0)$] between the states corresponding to equal occupations $c_n$. This occurs, for example, when $\hat{\rho}_{\mathrm{E}}(0)$ is a function of one of $\hat{H}_{i}$. The subspaces of equal $c_n$ are then subspaces of equal energy. The most natural physical situation when this occurs is when the environment thermalizes according to $\hat{H}_{i}$. This can happen when one of the qubit-environment interaction terms, say $\hat{V}_{0}$, vanishes (so that $\hat{H}_{0} \! =\! \hat{H}_{\mathrm{E}}$), and E is allowed to reach thermal equilibrium in the absence of the qubit. 
This is the case for quantum dot excitonic qubit (with $\ket{0}$ corresponding to the optical ground state) coupled to a bath of phonons \cite{Krummheuer_PRB02,Roszak_PLA06}. Interestingly, in qubits based on 
the nitrogen-vacancy center in diamond \cite{Doherty_PR13} it is possible to realize both $\hat{V}_{0} \! = \! 0$ and $\neq \! 0$ cases (by appropriately choosing two qubit levels out of three available states of electronic complex with total spin $1$), and observe distinct Q decoherence behavior in the two of them \cite{Zhao_PRL11,Huang_NC11}.
Alternatively, the qubit can be initialized in its lowest-energy state for a time long enough for E thermalize in its presence \cite{Shnirman_PS02}, then the state of the environment is a function
of the Hamiltonian $\hat{H}_{0}$ without the additional constraint
of $\hat{V}_{0}=0$ (i.e.~it reaches the thermal state defined by $\hat{H}_{0}$ before the qubit is rotated into a superposition state and its dephasing begins). One should note that when thermal states of E are considered, the amount of generated QEE (presumably correlated with the magnitude of change of chosen environmental observable) is expected to be proportional to $\beta$, the inverse temperature of E (since for $\beta \! =\! 0$ there is no QEE). This could be useful for thermometry of an environment for which $k_{\mathrm{B}}T$ is large compared to typical energy scale of $\hat{H}_{i}$, e.g.~for a nuclear bath coupled to a spin qubit \cite{Coish_pssb09,Cywinski_APPA11}.  

In conclusion, we have studied the generation of qubit-environment entanglement (QEE) via pure dephasing processes and specified the only three classes of situations (specified by the initial state of E
and relevant evolution operator which is derived from the Hamiltonian) when QEE will not be generated.
These are, the case when the initial density matrix of the environment is proportional to 
unity, the case when the relevant evolution operator cannot change the occupation of any of the eigenstates
of the density matrix (the random unitary evolution due to a quasi-static bath, essentially inhomogenoeous broadening of Q splitting in the context of pure dephasing), 
and a non-trivial mixture of the two cases which allows dynamical evolution within closed subspaces of equal occupation. 
Furthermore, we have shown that restricting the initial E states to a class which is very common in any realistic qubit-environment setup, enables the use of a very powerful tool to measure QEE, since the state of E will remain static throughout the evolution if no QEE is generated. Hence, the detection of any change of the state of E is then equivalent to the detection of entanglement. 

This work is supported by funds of Polish National Science Centre (NCN) under decision no.~DEC-2012/07/B/ST3/03616. \L. C. thanks Igor Bragar for assistance with numerical testing of hypotheses at an early stage of reseach, and Cezary {\'S}liwa for useful comments.

\appendix
\section*{Appendix}
\subsection{An example of evolution leading to entanglement of a qubit with maximally mixed environment}
Let us give a very simple example of a unitary qubit-environment operation (generated by a Hamiltonian more general than the one correspoding to pure dephasing of the qubit), which leads to creation of QEE at some time $t$ after initialization of a pure state of Q and a completely mixed state of E. We take a two-dimensional environment and take a unitary qubit-environment operation $\hat{U}_{\mathrm{C}}$ that transforms
a given qubit state $|\psi\rangle$ together with the states $|0\rangle,|1\rangle$ of a two-dimensional environment as follows: 
\begin{align}
\hat{U}_{\mathrm{C}}|\psi0\rangle & = |01\rangle \,\, , \\ 
\hat{U}_{\mathrm{C}}|\psi1\rangle & = (|00\rangle+|11\rangle)/\sqrt{2} \,\, ,\\
\hat{U}_{\mathrm{C}}|\psi_{\perp}0\rangle & = (|00\rangle-|11\rangle)/\sqrt{2} \,\, ,\\
\hat{U}_{\mathrm{C}}|\psi_{\perp}1\rangle & =|10\rangle \,\, ,
\end{align}
where $|ij\rangle$ denotes the state $|i\rangle_{Q}\otimes|j\rangle_{E}$, and $|\psi_{\perp}\rangle$ is orthogonal to $|\psi\rangle$. In the pure dephasing (PD) case the evolution is such that there exists a unitary operation on the qubit, that brings $\hat{U}_{\mathrm{C}}$ to block-diagonal form with respect to states of Q. It is straightforward to check that such an unitary operation does not exist in the above-described case.
Under the action of $\hat{U}_{\mathrm{C}}$ an initial state of the form $|\psi\rangle\langle\psi |\otimes\mathds{1}_{\mathrm{E}}/2$
is transformed into the state (in the qubit-environment basis $\{|00\rangle,|01\rangle,|10\rangle,|11\rangle\}$)
\begin{equation}
\hat{U}_{\mathrm{C}}|\psi\rangle\langle\psi |\otimes\mathds{1}_{\mathrm{E}}\hat{U}^{\dagger}_{\mathrm{C}}=
\left(
\begin{array}{cccc}
1/4&0&0&1/4\\
0&1/2&0&0\\
0&0&0&0\\
1/4&0&0&1/4
\end{array}
\right),
\end{equation}
which is entangled (the concurrence of this state is equal to 1/2). We have thus shown that there exist such initial states $|\psi\rangle$ of the qubit, and such Hamiltonians of the whole system, that lead to creation of entanglement at some time $t$ after initialization of the qubit and the two-dimensional environment in a separable state, with the environment being initially maximally mixed.

\end{document}